# Non-linear diffusion of negatively charged excitons in WSe$_2$ monolayer

D. Beret[1], L. Ren[1], C. Robert[1], L. Foussat[1], P. Renucci[1], D. Lagarde[1], A. Balocchi[1], T. Amand[1], B. Urbaszek[1], K. Watanabe[2,3], T. Taniguchi[2,3], X. Marie[1] and L. Lombez[1]

[1]Université de Toulouse, INSA-CNRS-UPS, LPCNO, 135 Av. Rangueil, 31077 Toulouse, France
[2]International Center for Materials Nanoerchitectonics, National Institute for Materials Science, 1-1 Namiki, Tsukuba 305-00044, Japan
[3]Research Center for Functional Materials, National Institute for Materials Science, 1-1 Namiki, Tsukuba 305-00044, Japan

**Abstract**

*We investigate the diffusion process of negatively charged excitons (trions) in WSe$_2$ transition metal dichalcogenide monolayer. We measure time-resolved photoluminescence spatial profiles of these excitonic complexes which exhibit a non-linear diffusion process with an effective negative diffusion behavior. Specifically, we examine the dynamics of the two negatively charged bright excitons (intervalley and intravalley trion) as well as the dark trion. The time evolution allows us to identify the interplay of different excitonic species: the trionic species appear after the neutral excitonic ones, consistent with a bimolecular formation mechanism. Using the experimental observations, we propose a phenomenological model suggesting the coexistence of two populations: a first one exhibiting a fast and efficient diffusion mechanism and a second one with a slower dynamics and a less efficient diffusion process. These two contributions could be attributed to hot and cold trion populations.*

The great recent interest in transition metal dichalcogenides (TMD) monolayers relies on their direct bandgaps and a strong light-matter interaction making these materials suitable for ultrathin optoelectronic devices [1]. The unusual optical and electrical properties of such monolayer semiconductors are related to the strong Coulomb interaction between electrons and holes, which open new ways to the development of devices based on excitonic species rather than free carriers, even at room temperature. Yet, a few experimental work have achieved the control of the excitonic transport by applying electrical gates [2] or by modifying the local strain [3,4]. The investigation of excitonic transport is therefore an important topic and several studies have already reported on the unassisted diffusion of several excitonic species [5–9]. Besides, in time and spatially resolved experiments, diffusion and recombination mechanisms (i.e. lifetimes) can be determined independently to better understand the transport beyond a simple measurement of the diffusion length.

Although efficient excitonic transport could be measured, it turns out that the rich family of excitonic species can make transport mechanism rather complex and nonlinear diffusion processes have been observed essentially linked to the interactions between the different excitonic complexes [10–16]. For instance, the classical picture of diffusion is questioned by the presence of nonequilibrium excitons which create a halo-like carrier profile induced by a hot population [10,13,17,18]. Such behavior might also be linked to the dielectric environment as the monolayer encapsulation impacts the transport mechanisms [19,20].

Another uncommon behavior is the prediction of an effective negative diffusion in TMD where a narrowing of the spatial exciton distribution coming from the interplay of excitonic species is expected [21]. Such behavior has been experimentally observed in several semiconducting materials such as perovskite [21], TMD [22] or organic semiconductor [23] at room temperature. In the two later cases, the co-existence of two populations (or two heat channels) governs the non-linear diffusion process.

Here, we study a $WSe_2$ monolayer which offers a rich variety of excitonic complexes. We use a state-of-the-art charge adjustable device and focus on the n-doped regime to probe the influence of electron doping on the transport properties. We therefore study negatively charged excitons (trions) and investigate bright and dark complexes at low temperature. Our study reveals a non-linear transport behavior where an effective negative diffusion is observed. A simple diffusion model evidences the co-existence of two populations that could be related to hot and cold trion populations. The interplay with the neutral excitons through bimolecular formation with resident electrons is also evidenced.

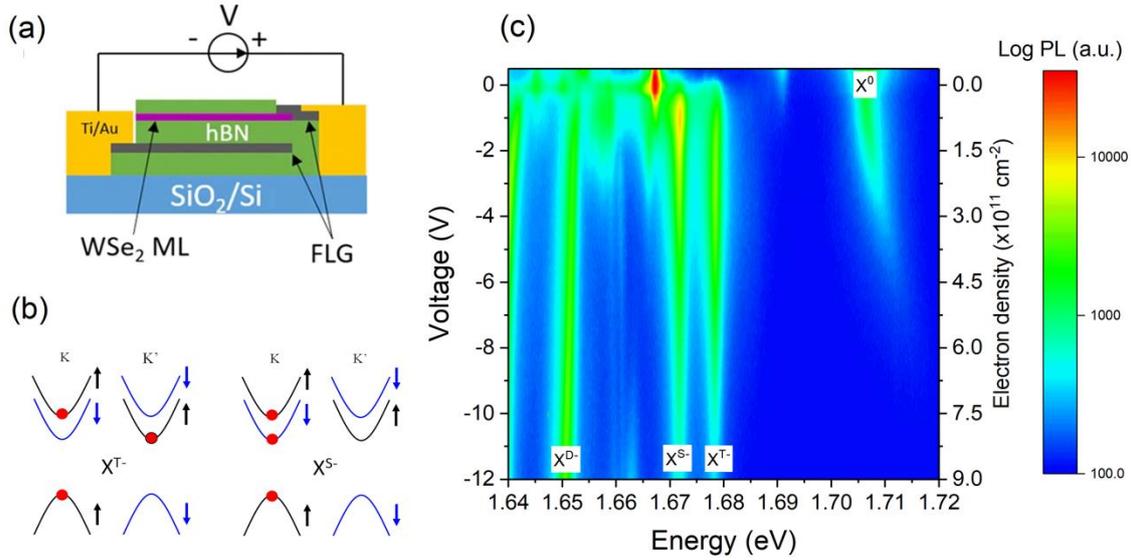

*Figure 1: (a) Scheme of the sample structure, ML=monolayer, FLG: few layer graphene (b) sketch of the two negative trion configurations : triplet $X^{T-}$ and singlet $X^{S-}$ (c) Colormap showing the evolution of the PL spectra with the bias (i.e.. electrostatic doping).*

**Experimental details**

We study a charge adjustable $WSe_2$ device which is sketched in Figure 1(a). Details of the sample fabrication can be found elsewhere [24]. An electric bias can be applied to tune the carrier concentration and we focus our analysis on the n-doped regime [24–26]. Typical PL spectra as a function of the electron doping concentration are displayed in Figure 1(c). Measurements are performed at 4K with a 5µW He:Ne laser excitation at $E_{ex}$=1.96eV. Several excitonic species are identified such as the bright neutral exciton $X^0$, the dark neutral exciton $X^D$, the bright negative trions (intervalley/triplet $X^{T-}$ and intravalley/singlet $X^{S-}$ see Figure 1(b)) and the negatively charged dark trion $X^{D-}$ in agreement with previous reports [25,27–32]. The encapsulation by h-BN layers offers state-of-the art quality noticeable by the narrow linewidth of the transition such as a 2.5 meV FWHM linewidth for the $X^0$ transition in the neutral regime [33].

In the following we mainly focus on the transport properties of the two bright trions (negatively charged exciton) $X^{T-}$ and $X^{S-}$. As sketched in Figure 1(b) the triplet $X^{T-}$ is composed of a bright exciton and an electron in the opposite valley whereas the singlet $X^{S-}$ is composed of a bright exciton and a resident electron in the same valley. The photoluminescence experiment is based on diffraction-limited laser excitation that induces a lateral diffusion of the photogenerated species [34]. The investigated excitonic complex is spectrally selected by adding angle tunable long and short pass interferential filters which transmit a spectral region as narrow as 6 meV (2.5 nm). We then used a Streak camera system with a temporal resolution of about 3 ps to record the time evolution of the PL spatial profile [15]. The Ti:Sa laser excitation is set at $E_{ex}$=1.79eV, 80MHz, 1.5ps pulse width and with a pulse energy density of about 1µJ/cm². The spatiotemporal information allows separating the transport and the recombination mechanisms [35]. The PL spatial profile is considered as gaussian $\forall t : I_{PL}(x) \sim \exp\left(-\frac{x^2}{w^2}\right)$ and we examine the squared width $w^2$ of the PL profile as a function of time w²(t) [35]. In case of linear broadening (i.e. classical diffusion process) an effective diffusion coefficient D can be extracted according to the relation w²(t) =4Dt [35]. However, this simple relation is not applicable if a non-linear behavior is observed.

We first describe the time dependence of the squared width of the PL profiles w² for the trion $X^{T-}$ displayed in Figure 2 for different electron doping densities. Note that similar observations are made for the trion $X^{S-}$ with a small difference on the dynamics as the two bright trions have slightly different lifetimes (see Supplemental Material) [36]. Three regimes can be identified. (I) Within a few picoseconds there is a fast broadening which would indicate an efficient diffusion process with an effective diffusion coefficient close to 1000 cm²/s. For higher electron doping densities, this first regime becomes less visible and the PL profile is already broader than the laser spot indicating a very fast diffusion (even shorter than our temporal resolution at higher doping density). In the next 20 picoseconds, a second regime (II) is visible with a clear reduction of the PL profile width, mimicking an effective negative diffusion process. The insets in Figure 2 represent the spatial profiles when the squared width w² is maximum and a few picoseconds later to better illustrate this second regime. (III) Finally, on a longer time scale, we observe a weak diffusion process with a weak dependence on the n-doping.

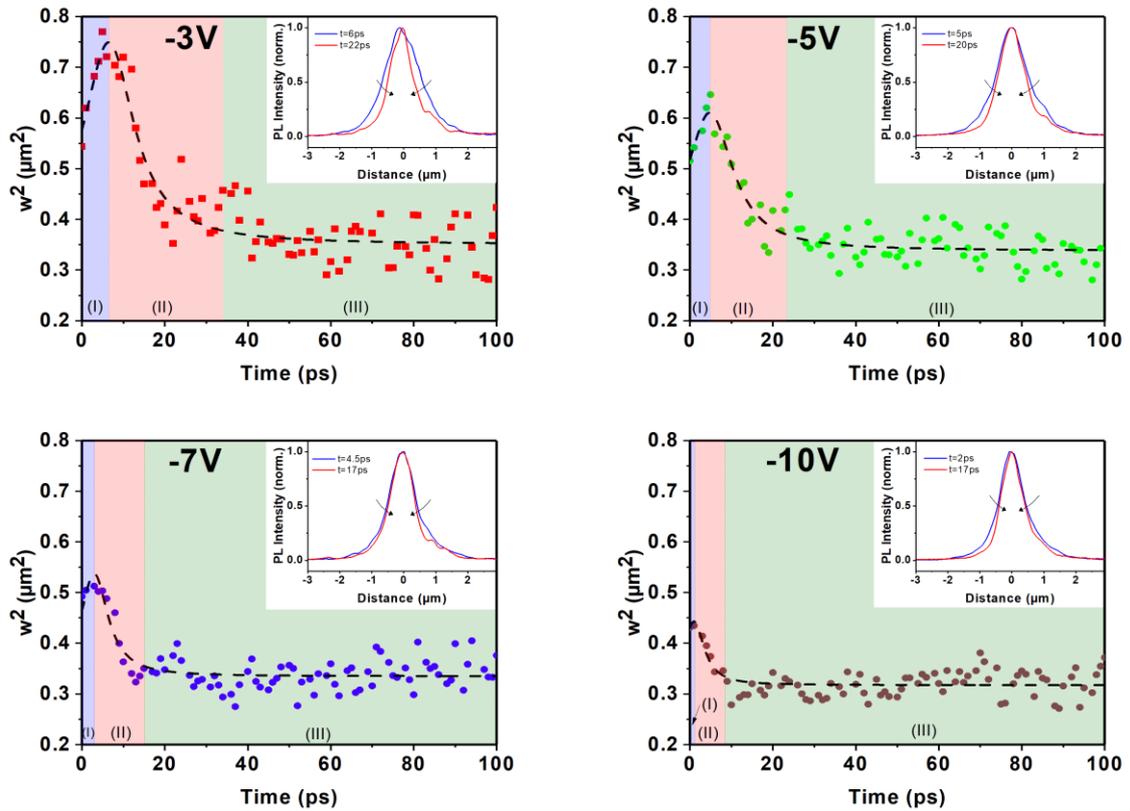

*Figure 2: Squared width of the PL spatial profile of the trion $X^{T-}$ as a function of time. The temporal evolution is displayed for different bias from -3V to -10V. The insets represent the spatial profiles at short time when the squared width w² is maximum and a few picoseconds later after the effective negative diffusion. Temporal regimes I (purple) II (red) and III (green) are indicated on each panel.*

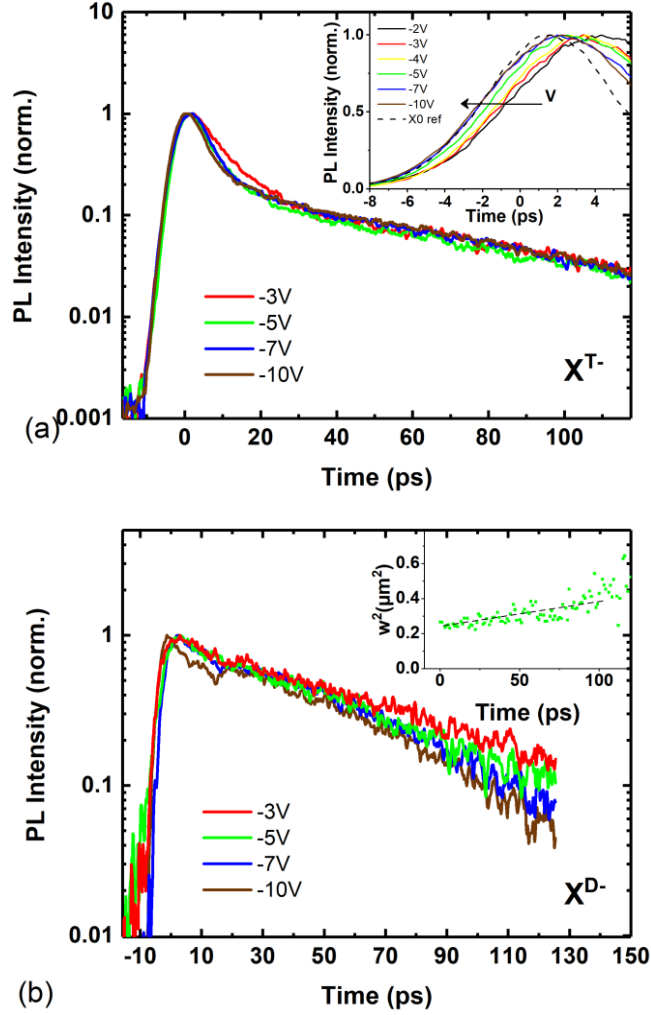

Figure 3: (a) TRPL dynamics for $X^{T-}$ at different bias. Inset : TRPL dynamics for the first picoseconds for $X^{T-}$ ; $X^0$ is shown as the time reference. (b) TRPL dynamics for $X^{D-}$ at different bias. Inset : squared width PL profile as a function of time for V=-3V.

We now describe time resolved photoluminescence data and we will focus on the interplay of the excitonic species. Figure 3(a) shows the spatially integrated time evolution of the PL intensity (TRPL) for different bias and for the excitonic complexes $X^{T-}$, while Figure 3(b) shows the TRPL intensity dependences for $X^{D-}$. The TRPL dynamics of the dark complex is longer than the bright complexes [24,25,37] and shows a rather mono-exponential behavior as compared to the bright complexes. Another difference is the temporal evolution of the squared width w² for $X^{D-}$ which show a linear dependence as it is seen in the inset of Figure 3(b) and in the supplemental material for all the voltages.

The TRPL of $X^{T-}$ exhibits a fast decay followed by a slow one; the fast dynamics becomes progressively shorter with increasing the electron concentration and mostly corresponds to the second regime (II) mentioned above, see Figure 2. If we now look at the first picoseconds (see inset Figure 3(a)) we observe that negatively charged excitonic complexes appear with a clear delay after the emission of the neutral exciton. Note that all negatively charged excitonic complexes ($X^{S-}$,$X^{T-}$, $X^{D-}$ not shown here) appear after the neutral exciton $X^0$. We find that the higher the doping, the shorter the delay times. This observation is consistent with a bimolecular formation process of the trionic species between the neutral exciton and the resident electrons.

A trimolecular formation process seems less probable as the TRPL signal of the trion in the early stage is negligible as compare to the $X^0$. Moreover, the laser excitation energy is below the bandgap energy [38]; it would suggest a geminate formation of the $X^0$ excitonic species and make the trimolecular formation of the trions from free charges very unlikely.

**Discussion**

We now discuss the origin of the non-linear diffusion regime of the bright negatively charged excitons. We focus once again on the triplet $X^{T-}$ but similar conclusions can be applied to the singlet $X^{S-}$. To model the effective negative diffusion, we considering one single population is insufficient. Indeed, the solution of the classical diffusion equation cannot render such behavior, even considering a time dependence of the diffusion coefficient. For instance, one can model a reduction of the diffusion coefficient when the carrier concentration is decreasing as it becomes more sensitive to potential fluctuations or defects [39]. But even in that scenario, the temporal evolution of w² is slowed down but never decreases. A time dependence of the diffusion coefficient modifies the diffusion efficiency but always in a monotonic way. A phenomenological explanation might come from the contribution of two populations that cannot be spectrally distinguished within the ~6meV width of our spectral selection resolution. The first having a *fast* diffusion mechanism combined with a shorter effective lifetime and a second presenting a *slow* diffusion process and a longer effective lifetime. In this case, the first population contributes to the rapid increase of w² while vanishing quickly to make room for the second contribution having a weak role on the diffusion process.

In order to qualitatively model the non-linear diffusion with the effective negative diffusion, we consider two populations $n_{X^{T1}}$ and $n_{X^{T2}}$ contributing to the PL intensity of $X^{T-}$. These populations are set from a neutral exciton density $n_{X^0}$ via bimolecular formation with the electron reservoir $n_{e^-}$ (see scheme in Figure 4(a)). The two populations arise from the solution of the simple diffusion equation in polar coordinates when considering similar recombination lifetimes $\tau_{X^{T1}} = \tau_{X^{T1}}$ but different diffusion coefficients $D_{X^{T1}} \neq D_{X^{T2}}$. Note that although $\tau_{X^{T1}} = \tau_{X^{T2}}$ the effective lifetime of the two populations is different (see Figure 4(a)) since the spatial dependence is linked to the temporal dependence. The set of partial differential equations writes as :

$$\frac{dn_{X^0}}{dt} = G + \vec{\nabla_r}.\left[D_{X^0}\vec{\nabla_r}n_{X^0}\right] - \frac{n_{X^0}}{\tau_{X^0}} - B_1 n_{X^0} n_{e^-} - B_2 n_{X^0} n_{e^-}$$

$$\frac{dn_{X^{T1}}}{dt} = \vec{\nabla_r}\left[D_{X^{T1}}\vec{\nabla_r}n_{X^{T1}}\right] - \frac{n_{X^{T1}}}{\tau_{X^{T1}}} + B_1 n_{X^0} n_{e^-}$$

$$\frac{dn_{X^{T2}}}{dt} = \vec{\nabla_r}\left[D_{X^{T2}}\vec{\nabla_r}n_{X^{T2}}\right] - \frac{n_{X^{T2}}}{\tau_{X^{T2}}} + B_2 n_{X^0} n_{e^-}$$

Solving these equations yields a space-time evolution of the trion PL intensity $I_{PL}(r,t) = {n_{X^{T1}}}/{\tau_{X^{T1}}} + {n_{X^{T2}}}/{\tau_{X^{T2}}}$. We then extract the temporal dependence of the spatial PL profile with the same method used to treat the experimental data. The modeling of w²(t) for each individual trion contribution as well their sum (i.e. the PL intensity) is presented in Figure 4(b). A qualitative agreement is found with our experimental data. To set the lifetime values we refer to the ones extracted from our TRPL data which are similar to those found in the literature, $\tau_{X^0} = 1ps$ and $\tau_{X^{T1}} = \tau_{X^{T2}} = 60ps$ [24]. The bimolecular recombination coefficients are $B_1$=0.9x10$^{-11}$ cm²/ps and $B_2$=0.1x10$^{-11}$ cm²/ps, also in the same order of magnitude to the one we used previously to model experimental data from the same sample [24]. The diffusion coefficients $D_{X^0} = D_{X^{T2}}$ =1cm²/s are typical for excitonic species [19] while $D_{X^{T1}}$=1000cm²/s is much larger. However, this high value can also be evaluated by considering a linear diffusion process in the first picoseconds (region I) with the relation w² =4Dt. In the literature, such high values are also found, for example in WSe$_2$, when measuring the hot exciton diffusion at room temperature [22].

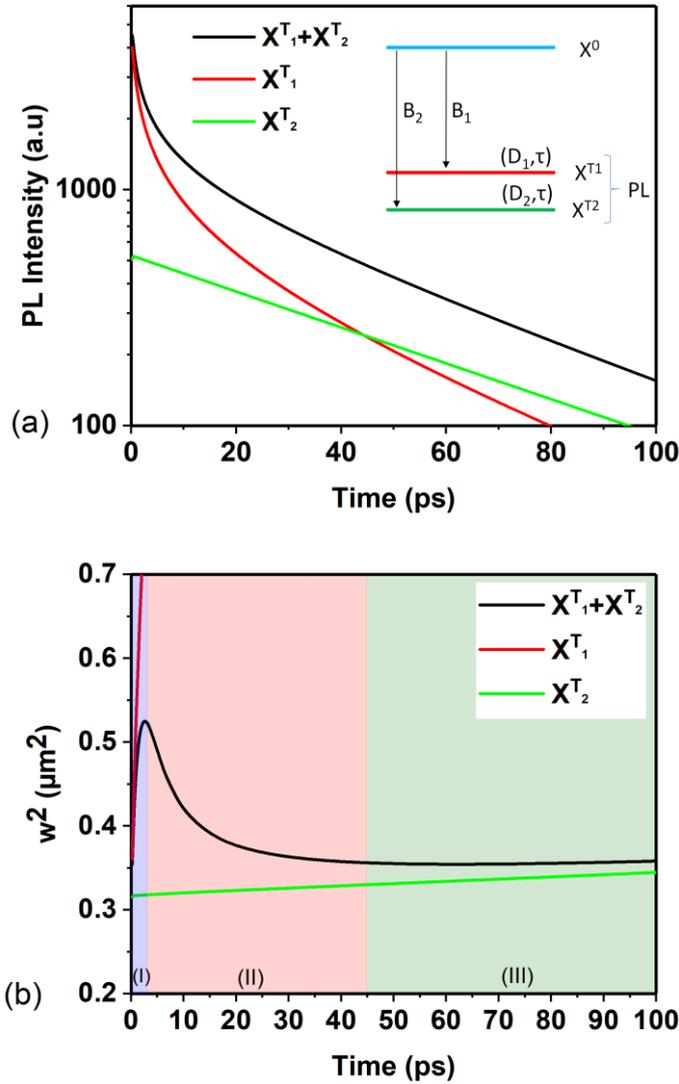

Figure 4: (a) TRPL simulated signal for $I_{X^{T1}}$, $I_{X^{T2}}$ and $I_{PL} = I_{X^{T2}} + I_{X^{T1}}$ in logarithmic scale (b) Simulated temporal evolution of the squared width w² of the PL spatial profiles, see text.

To better illustrate the role of the two populations in the non-linear diffusion process, we plot in Figure 4(a) the simulated TRPL signals I$_{PL}$ for V=-3V as well as the contribution of the populations $I_{PL} = I_{X^{T2}} + I_{X^{T1}}$ . Figure 4(b) represents the temporal evolution of w² and the three regimes we observed experimentally have been colored. As in the experimental results, the first regime (region I) shows a positive and rapid classical diffusion. It is then followed by an effective negative diffusion behavior (region II). The change between the two regimes occurs when the contribution of the PL intensity of X$^{T1}$ becomes small enough to observe the PL intensity of X$^{T2}$. Finally, a small variation of w² is observed (region III) after a certain delay where the contribution of X$^{T1}$ becomes negligible compared to X$^{T2}$. Note that if we could detect the individual contribution of the two populations, we would obtain the two squared width dependence displayed in Figure 4(b) : in red from the PL signal of $X^{T1}$ and in green from the PL signal of $X^{T2}$. The fast contribution would give an effective coefficient of about 450cm²/s if we use the linear relation w²=4Dt while we found an effective diffusion coefficient of about 3 cm²/s by considering the slow contribution of $X^{T2}$.

Although the three-population model reproduces the non-linear behavior with relatively good agreement, the influence of the electron doping cannot be reproduced. This model is quite insensitive to the variation of $n_{e-}$. Indeed, a change of the latter parameter has a weak influence on the ratio $n_{X^{T1}}(t)/n_{X^{T2}}(t)$ which is responsible for the change of w²(t) (See Supplemental Material for details). In order to improve the model, we introduce two neutral excitonic populations : a hot and a cold exciton population, as already considered for the interpretation of diffusion process in a previous study [39]. In this case by splitting the first equation of the population $n_{X^0}$ in two parts (See SI), the voltage dependence is reproduced as seen in Figure 5.

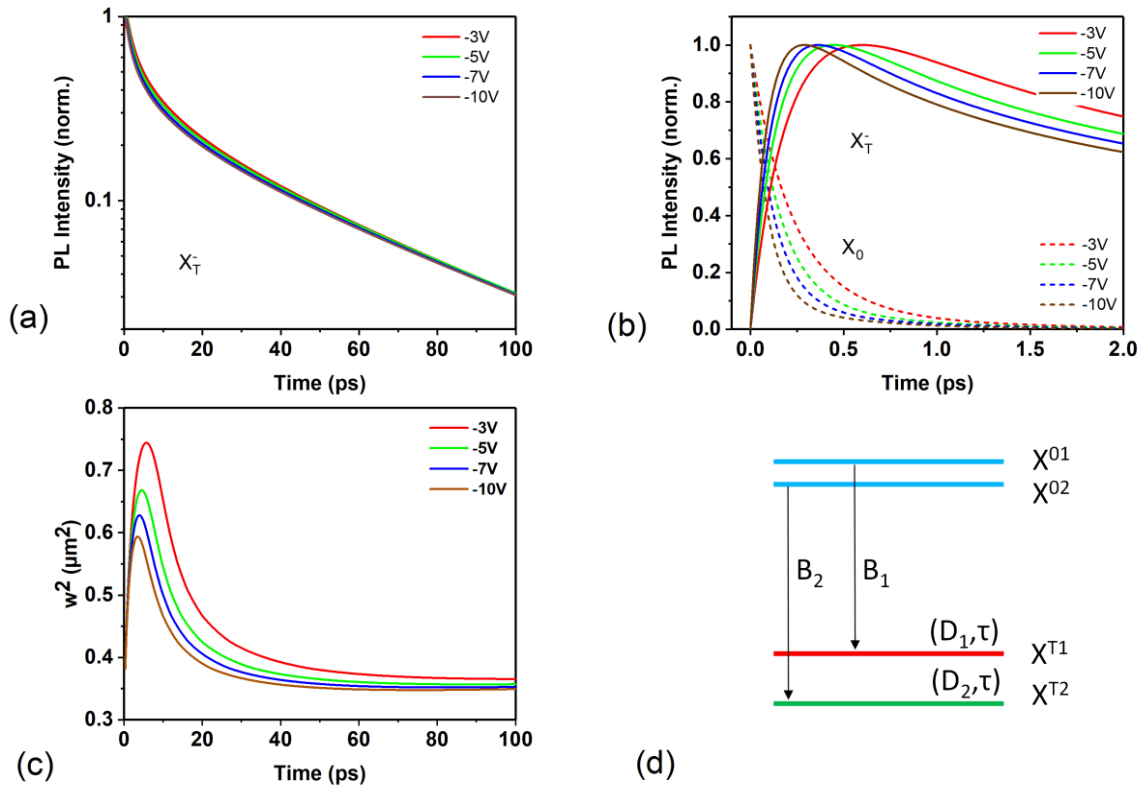

Figure 5: (a) TRPL simulated signal for $X_{T^-}$ for different bias (i.e. electron doping density) (b) TRPL simulated signal in the first picoseconds for $X_0$ and for $X_{T^-}$ different bias (i.e. electron doping density). (c) Simulated temporal evolution of the PL profile with w² for different bias (d) Sketch of the energy levels we use for the model including two trion populations $X^{T1}$ and $X^{T2}$ as well as two neutral exciton populations $X^{01}$ and $X^{02}$.

As the electron doping is increased, the fast contribution giving the increase of the PL profile squared width w² decreases and appears earlier; in other words, the slow contribution becomes dominant (See Figure 5(c)). TRPL data are also reproduced with a good agreement as observed in Figure 5(a) where a slight variation with the bias is seen at short time. In particular, the time delay between the neutral exciton $X^0$ PL emission and the trion PL emission is also found to be reduced as we increase the electron doping concentration (see Figure 5(b)); this in agreement with the experimental observation (inset Figure 3(a)).

The origin of the two trions populations is still unclear. As mentioned earlier, it is unlikely that the experimental results are related to defects or disorder as a unique diffusion coefficient cannot reproduce the experimental observation. This is also in line with the narrow PL linewidths of the excitonic species which suggest a high-quality sample. One possibility would be the presence of hot and cold trion populations that we cannot distinguish within the 6meV of our spectral window. Hot population of trionic species has been shown to slightly increase the PL spectral width at low energy [40] and its dynamics can correspond to what is theoretically found in the literature on $WSe_2$ [41], especially having a hot population lifetime of a few ps. This hypothesis is also supported by the experimental observation of excitonic complexes at 300K in perovskite materials [21] and $WS_2$ [22]. As mentioned in the introduction, theorical investigation on effective negative diffusion might also imply thermalization mechanisms [42]. Finally, we point out that no effective negative diffusion is seen on the negatively charged dark trion species. This might be explained by the relatively long lifetime where efficient thermalization might occur before recombination. The mono-exponential TRPL signal of this excitonic complex could also be an indication of the relaxation of a single population as compared to the negatively charged bright excitons for which a rather bi-exponential decay is observed. The diffusion coefficient of $X^{D-}$ is slightly reduced with the electron doping (see SI) which would be consistent with a reduction of the thermalization while increasing the cold electronic reservoir. Overall the experimental data and the modeling would suggest efficient transport mechanisms assisted by hot population. This is visible if the effective lifetime of the excitonic complex is sufficiently small compared to the thermalization time. To confirm such hypothesis, a pump-probe experiment combining picosecond time resolution and meV spectral resolution could be performed in the future.

**Conclusion**

We have investigated the transport properties of negatively charged excitonic complexes in $WSe_2$ monolayers. A non-linear diffusion process is revealed for the two bright trions where an effective negative diffusion is evidenced. A phenomenological model highlights the contribution of two populations with respectively a fast and efficient diffusion mechanism combined with a slow but inefficient one. The coexistence of these two populations might be present for the neutral exciton as well and could originate from a hot and cold populations. As a general trend, hot populations would have a large diffusion coefficient before they thermalize. Although this hypothesis might require further experimental studies it points out that efficient transport mechanism in TMD materials might be tuned by controlling the thermalization rate.

**Acknowledgement**

This study has been partially supported through the EUR grant NanoX n° ANR-17-EURE-0009 in the framework of the « Programme des Investissements d'Avenir». We thank F. Cadiz for fruitful discussions. This work was also supported by Agence Nationale de la Recherche funding ANR SIZMO2D, ANR ICEMAN, ANR ATOEMS, ANR IXTASE

# SUPPLEMENTAL MATERIAL

## Non-linear diffusion of negatively charged excitons in WSe$_2$ monolayer

D. Beret[1], L. Ren[1], C. Robert[1], L. Foussat[1], P. Renucci[1], D. Lagarde[1], A. Balocchi[1], T. Amand[1], B. Urbaszek[1], K. Watanabe[2,3], T. Taniguchi[2,3], X. Marie[1] and L. Lombez[1]

[1]Université de Toulouse, INSA-CNRS-UPS, LPCNO, 135 Av. Rangueil, 31077 Toulouse, France
[2]International Center for Materials Nanoerchitectonics, National Institute for Materials Science, 1-1 Namiki, Tsukuba 305-00044, Japan
[3]Research Center for Functional Materials, National Institute for Materials Science, 1-1 Namiki, Tsukuba 305-00044, Japan

## I. Intravalley negatively charged exciton (singlet) $X^{S-}$

We present below in Figure 6 the time dependence of the squared width w² for the singet $X^{S-}$ (see main text for description). Similar observations as for the triplet are made. When we increase the bias the effective negative diffusion comes sooner and the maximum of w² is smaller.

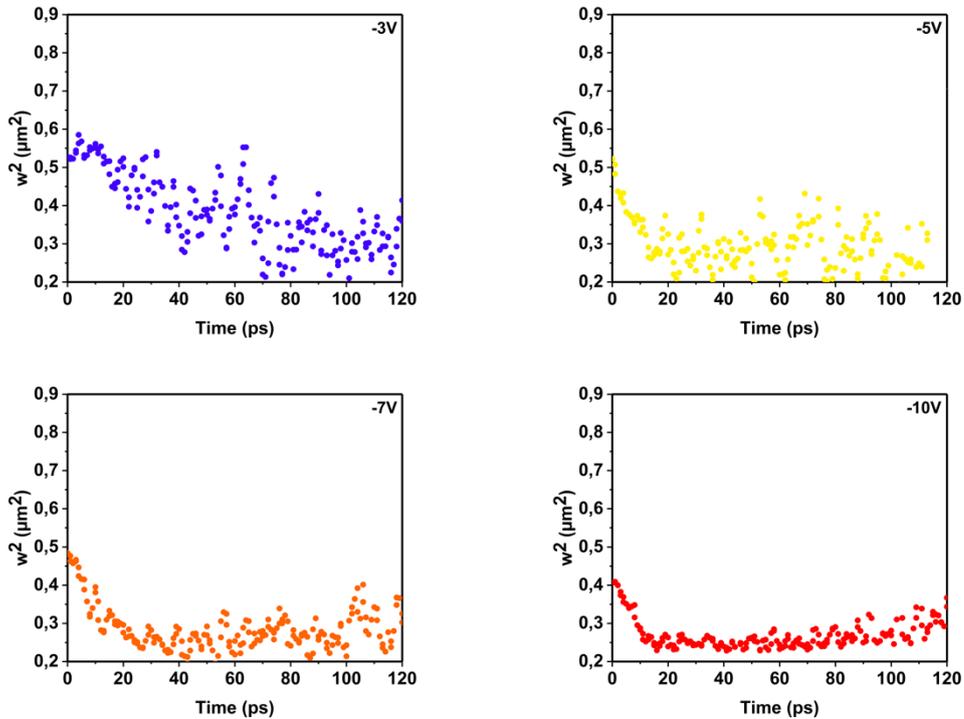

*Figure 6 : Squared width of the PL spatial profile as a function of time for the singlet $X^{S-}$. The temporal evolution is displayed for different bias from -3V to -10V.*

## II. Negatively charged dark trion $X^{D-}$

We present in Figure 7 the temporal evolution of the squared width PL profiles for the negatively charged dark exciton. We notice a slight reduction of the diffusion coefficient by increasing the bias (i.e. electron doping concentration).

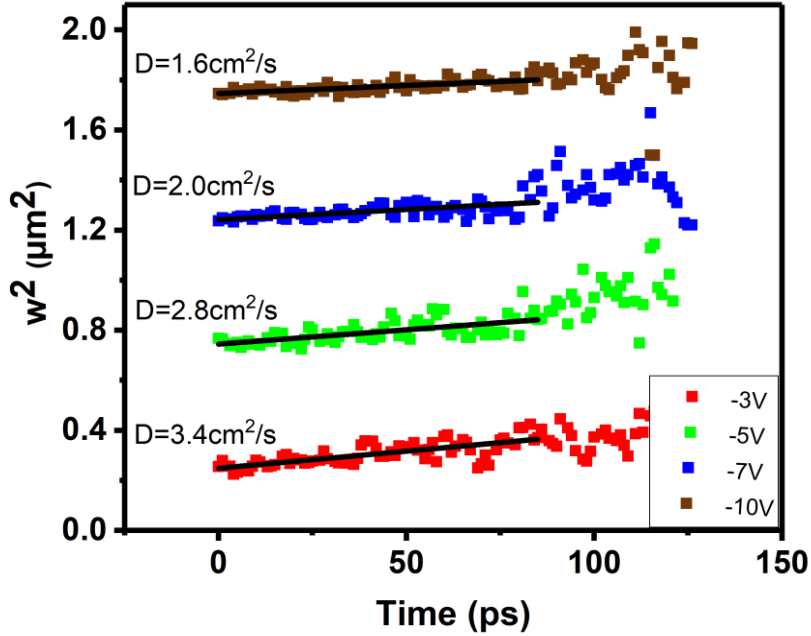

*Figure 7 : Squared width of the PL spatial profile as a function of time for the negatively charged dark exciton $X^{D-}$. The temporal evolution is displayed for different bias from -3V to -10V. The curves are shifted vertically for clarity, the slope remaining valid to estimate the diffusion coefficient.*

### III. 3-population model vs. 4-population model

The 3-population model is described in the main text, we present here the model with 4-population : 2 excitons $n_{X^{01}}$ and $n_{X^{02}}$ and 2 trions $n_{X^{T1}}$ and $n_{X^{T2}}$. The set of diffusion equations writes as :

$$\frac{dn_{X^{01}}}{dt} = G_1 + \vec{\nabla_r}.\left[D_{X^{01}}\vec{\nabla_r}n_{X^{01}}\right] - \frac{n_{X^{01}}}{\tau_{X^0}} - B_1 n_{X^{01}} n_{e^-}$$

$$\frac{dn_{X^{02}}}{dt} = G_2 + \vec{\nabla_r}.\left[D_{X^{02}}\vec{\nabla_r}n_{X^{02}}\right] - \frac{n_{X^{02}}}{\tau_{X^0}} - B_2 n_{X^{02}} n_{e^-}$$

$$\frac{dn_{X^{T1}}}{dt} = \vec{\nabla_r}\left[D_{X^{T1}}\vec{\nabla_r}n_{X^{T1}}\right] - \frac{n_{X^{T1}}}{\tau_{X^{T1}}} + B_1 n_{X^{01}} n_{e^-}$$

$$\frac{dn_{X^{T2}}}{dt} = \vec{\nabla_r}\left[D_{X^{T2}}\vec{\nabla_r}n_{X^{T2}}\right] - \frac{n_{X^{T2}}}{\tau_{X^{T2}}} + B_2 n_{X^{02}} n_{e^-}$$

For the sake of simplicity we did not consider thermalization rate and, as mentioned in the main text, we use our TRPL data (which match with data in literature) for setting the different lifetimes and Bimolecular recombination coefficients close to the one use in previous study [24]. We use $\tau_{X^0} = 1ps$ and $\tau_{X^{T1}} = \tau_{X^{T2}} = 60ps$. The bimolecular recombination coefficients are $B_1$=0.9x10$^{-11}$ cm²/ps and $B_2$=0.1x10$^{-11}$ cm²/ps. The diffusion coefficients $D_{X^{02}} = D_{X^{T2}}$=1cm²/s are typical for excitonic species, which could be ascribed to the cold excitonic species [19] while we set $D_{X^{01}} = D_{X^{T1}}$=1000cm²/s for the efficient diffusion mechanisms of complexes that could be ascribed to hot species.

Solving those equations yields a space-time evolution of the trion PL intensity $I_{PL}(r,t) = n_{X^{T1}}/\tau_{X^{T1}} + n_{X^{T2}}/\tau_{X^{T2}}$ where we extract the TRPL data by looking at the spatial integration of the PL signal as well as the squared width PL profile w² assuming a Gaussian profile.

The comparison of the two model for different voltages is presented in Figure 8 where Fig8(a) and (c) show the time dependence of the squared width w² for the 3-population model and the 4-population model respectively; while Fig.8(b) and (d) show the time dependence of the ratio $n_{X^{T1}}/n_{X^{T2}}$ which affect the time dependence of w².

In the case of the 3 populations model we see a weak dependence on the voltage (i.e. on the electron doping concentration). The 4-population model also offers a better agreement with the experimental results.

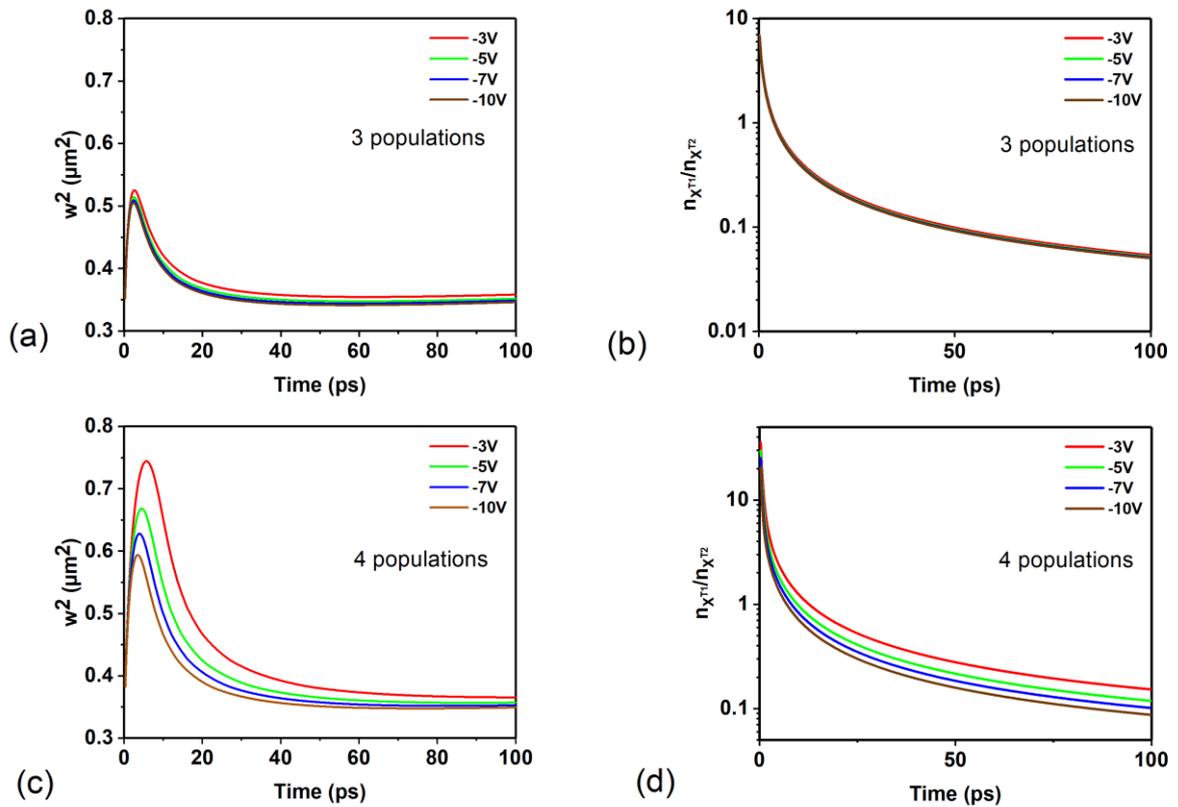

*Figure 8 : (a,c) time evolution of the squared PL profile for the 3-population model and the 4-population model respectively (b,d) time evolution of the ratio of the two trionic populations ratio $n_{X^{T1}}/n_{X^{T2}}$ as a profile for the 3-population model and the 4-population model respectively.*